\documentclass[noinfo,nocrop]{ipart_v1}

\usepackage[T1]{fontenc}
\usepackage[latin9]{inputenc}
\usepackage[english]{babel}
\usepackage{amsthm}
\usepackage{amssymb}
\usepackage{esint}
\makeatletter

\providecommand{\tabularnewline}{\\}

  \theoremstyle{definition}
  \newtheorem*{example*}{\protect\examplename}

\usepackage{tikz}
\usetikzlibrary{matrix,arrows,positioning}

\usepackage{import}

\newcommand{\Arg}{\mathrm{Arg}}

\makeatother

  \providecommand{\examplename}{Example}

\begin{document}

\title{Open Gromov-Witten Invariants from the Augmentation Polynomial}

\author{Matthew Mahowald}
\begin{abstract}
A conjecture of Aganagic and Vafa \cite{AV12} relates the open Gromov-Witten
theory of $X=\mathcal{O}_{\mathbb{P}^{1}}(-1,-1)$ to the augmentation
polynomial of Legendrian contact homology. We describe how to use
this conjecture to compute genus zero, one boundary component open
Gromov-Witten invariants for Lagrangian submanifolds $L_{K}\subset X$
obtained from the conormal bundles of knots $K\subset S^{3}$. This
computation is then performed for two non-toric examples (the figure-eight
and three-twist knots). For $(r,s)$ torus knots, the open Gromov-Witten
invariants can also be computed using Atiyah-Bott localization. Using
this result for the unknot and the $(3,2)$ torus knot, we show that
the augmentation polynomial can be derived from these open Gromov-Witten
invariants. 
\end{abstract}
\maketitle

\section{Introduction}

Gromov-Witten theory has famously benefited from its connections with
string dualities, first with mirror symmetry \cite{COGP91,Witten92},
and more recently, large $N$ duality \cite{GV99,OV00}. Beginning
with \cite{Kontsevich95}, for toric manifolds, Gromov-Witten invariants
associated to maps of closed surfaces have also been systematically
computed using localization \cite{GP99,CKYZ99,KZ99}. Although the
analogous constructions in open Gromov-Witten theory are not rigorously
defined in general, many of the same computational tools (such as
mirror symmetry and Atiyah-Bott localization) can still be applied.
In contrast to the closed theory, open Gromov-Witten theory also possesses
many direct relationships with knot theory. Large $N$ duality relates
Chern-Simons theory on $S^{3}$ to Gromov-Witten theory on $X=\mathcal{O}_{\mathbb{P}^{1}}(-1,-1)$
via the conifold transition. Wilson loops in Chern-Simons theory on
$S^{3}$ are also related to the HOMFLY polynomials of knots $K\subset S^{3}$
\cite{Witten89}. This relationship equates the colored HOMFLY polynomials
of knots $K$ with generating functions for open Gromov-Witten invariants
of $X$ with Lagrangian boundary $L_{K}$ obtained from the conormal
bundle $N_{K}^{*}\subset T^{*}S^{3}$, and has been checked for torus
knots in \cite{DSV13}. 

Recently, it has also been suggested that open Gromov-Witten theory
is related to another type of knot invariant arising in Legendrian
contact homology \cite{AV12,AENV14}. For a knot $K\subset S^{3}$,
Legendrian contact homology associates a dga $\mathcal{A}\left(\Lambda_{K}\right)$
to its unit conormal bundle $\Lambda_{K}\subset U^{*}S^{3}$. The
unit conormal bundle $\Lambda_{K}\approx T^{2}$ is a Legendrian submanifold
of the unit cotangent bundle $U^{*}S^{3}$ (a contact manifold), and
the differential on $\mathcal{A}\left(\Lambda_{K}\right)$ is obtained,
roughly speaking, from counts of maps of holomorphic disks to $\Lambda_{K}\times\mathbb{R}$.
An augmentation of $\mathcal{A}\left(\Lambda_{K}\right)$ is a dga
map $\epsilon:\mathcal{A}\left(\Lambda_{K}\right)\rightarrow\mathbb{C}$,
where $\mathbb{C}$ is interpreted as a dga with trivial differential.
The moduli space of such augmentations is described by an equation
$A_{K}\left(x,p,Q\right)=0$, where $x$, $p$, and $Q$ are generators
for $H_{2}(U^{*}S^{3},\Lambda_{K})$. $A_{K}$ is called the augmentation
polynomial of the knot $K$. (More detailed accounts of Legendrian
contact homology can be found in \cite{Ng12} and \cite{AENV14}). 

In \cite{AV12}, Aganagic and Vafa conjecture that the moduli space
of open Gromov-Witten invariants on $(X,L_{K})$ is encoded by the
augmentation polynomial $A_{K}$. The goal of this paper is to use
this conjecture to compute Gromov-Witten invariants and augmentation
polynomials. Crucially, the augmentation polynomial can be computed
for non-toric knots, and hence, this method can be applied in scenarios
where Atiyah-Bott localization cannot be used. On the other hand,
for $(r,s)$ torus knots, open Gromov-Witten invariants are known
from localization, and this data provides another means of obtaining
the augmentation polynomial of $K$.

\subsection{Organization of the paper}

This paper is organized in the following way. Section~\ref{sec:MS-and-Aug-poly}
reviews mirror symmetry for open Gromov-Witten invariants, and describes
Aganagic and Vafa's conjecture relating mirror symmetry and the augmentation
polynomial. Section~\ref{sec:Non-Toric-Examples} applies this conjecture
to compute open Gromov-Witten invariants in two non-toric examples.
Finally, Section~\ref{sec:Recovering-the-aug-poly} performs the
reverse of this computation: the open Gromov-Witten invariants associated
to torus knots are used to recover the corresponding augmentation
polynomials. (Note that for $(r,s)$ torus knots, open Gromov-Witten
invariants in framing $rs$ have been computed directly via localization
in \cite{DSV13}).

\section{Open String Mirror Symmetry and the Augmentation Polynomial\label{sec:MS-and-Aug-poly}}

Recently, new developments in knot theory and open topological string
theory have uncovered connections between open Gromov-Witten theory
and knot theory \cite{AV12,AENV14,DSV13,OV00}. The subject of interest
in this note is open Gromov-Witten theory for Lagrangian submanifolds
of $X=\mathcal{O}_{\mathbb{P}^{1}}(-1,-1)$. Let $\Sigma$ be a genus-zero
Riemann surface with one boundary component. Denote by $K_{d,w}$
the open Gromov-Witten invariant associated to a stable map $f:\Sigma\rightarrow X$
with Lagrangian boundary conditions on a Lagrangian submanifold $L\subset X$,
where $d=\left[f\left(\Sigma\right)\right]\in H_{2}\left(X,L\right)$
and $w=\left[f\left(\partial\Sigma\right)\right]\in H_{1}\left(L\right)$.
This section describes a technique for computing $K_{d,w}$ using
a mirror symmetry conjecture \cite{AV12,AENV14}. 

\begin{figure}
\begin{center}
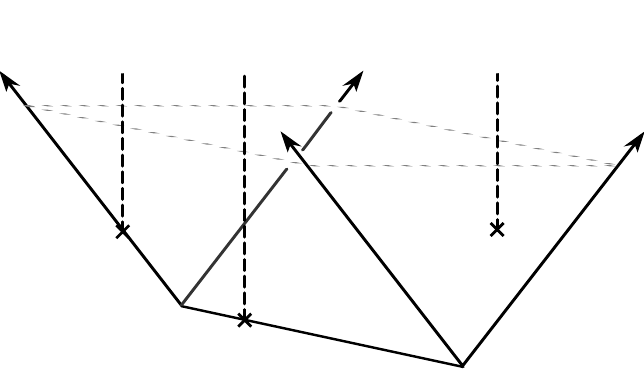
\end{center}

\caption{\label{fig:moment-polytope}The moment polytope and special Lagrangian
fibers of $X$.}

\begin{minipage}[t]{1\columnwidth}%
The moment polytope of $X=\mathcal{O}_{\mathbb{P}^{1}}(-1,-1)$ is
its image $\pi\left(X\right)\subset\mathbb{R}^{4}$ under the moment
map $\pi\left(z_{i}\right)=\left|z_{i}\right|^{2}$. The images of
the special Lagrangian fibers of $X$ in the moment polytope are vertical
lines, which can intersect $X$ three ways: along the base $\mathbb{P}^{1}$
(type I), along an exterior leg of the polytope (type II), or on a
face (type III). Lagrangian fibers of type III have topology $T^{2}\times\mathbb{R}$,
corresponding to $c_{1}\neq0$ and $c_{2}\neq0$ in (\ref{eq:sLag-fibers}).
Fibers of type I and II have topology $S^{1}\times\mathbb{R}^{2}$,
corresponding to either $c_{2}=0$ or $c_{1}=0$, respectively. %
\end{minipage}
\end{figure}

$X$ can be obtained by symplectic reduction on $\mathbb{C}^{4}$.
Let $(z_{1},z_{2},z_{3},z_{4})$ be coordinates for $\mathbb{C}^{4}$,
and let $S^{1}$ act on $\mathbb{C}^{4}$ with weights $(1,1,-1,-1)$;
then 
\begin{equation}
X\cong\left\{ \left|z_{1}\right|^{2}+\left|z_{2}\right|^{2}-\left|z_{3}\right|^{2}-\left|z_{4}\right|^{2}=r\right\} /S^{1},\label{eq:defn-of-X}
\end{equation}
where $r\in\mathbb{R}_{>0}$. The coordinates on the base $\mathbb{P}^{1}$
are $z_{1}$ and $z_{2}$, and the $z_{3}$, $z_{4}$ coordinates
parametrize the fiber. According to the Strominger-Yau-Zaslow conjecture
for noncompact $X$ \cite{SYZ96,AV00}, $X$ is a special Lagrangian
fibration over a base $B\cong\mathbb{R}^{3}$, with generic fibers
$L\cong T^{2}\times\mathbb{R}$. In these coordinates, the base $B$
and the special Lagrangian fibers $L$ are easy to describe. The base
$B$ is the image of $X$ under the moment map $z_{i}\mapsto\left|z_{i}\right|^{2}$,
and the fibers $L$ are given by the equations 
\begin{equation}
\begin{aligned}\left|z_{2}\right|^{2}-\left|z_{4}\right|^{2} & =c_{1},\\
\left|z_{3}\right|^{2}-\left|z_{4}\right|^{2} & =c_{2},\\
\Arg\left(z_{1}z_{2}z_{3}z_{4}\right) & =0,
\end{aligned}
\label{eq:sLag-fibers}
\end{equation}
where $c_{1},c_{2}\in\mathbb{R}$. For generic values of $c_{1}$,
$c_{2}$, $L$ has topology $T^{2}\times\mathbb{R}$; however, as
either $c_{1}\rightarrow0$ or $c_{2}\rightarrow0$, the topology
of the fibers degenerates to two copies of $S^{1}\times\mathbb{R}^{2}$.
This critical locus along which the fibers degenerate forms a trivalent
graph in $B$, corresponding to the ``edges'' of the moment polytope.
The moment polytope and special Lagrangian fibers are depicted in
Figure~\ref{fig:moment-polytope}.

\subsection{The Mirror of $\mathcal{O}_{\mathbb{P}^{1}}(-1,-1)$\label{sub:standard-mirror}}

The construction of \cite{HV00,AV00} gives the mirror manifold $\hat{X}$
to $X$ in terms of a dual Landau-Ginsburg theory. For $X=\mathcal{O}_{\mathbb{P}^{1}}(-1,-1)$,
the mirror equation is 
\[
\hat{X}=\left\{ uv=y_{2}+y_{3}+y_{4}+\frac{y_{3}y_{4}}{y_{2}}e^{-t}\right\} /\mathbb{C}^{*},
\]
where $u,v\in\mathbb{C}$, $y_{j}\in\mathbb{C}^{*}$, $t=r+i\theta$
is the complexified K\"{a}hler parameter, and the $\mathbb{C}^{*}$
acts diagonally on the $y_{j}$'s. The SYZ conjecture asserts that
the mirror $\hat{X}$ is the moduli of the special Lagrangian fibers
of $X$. The choice of coordinate patch for the $y_{j}$'s determines
which ``phase'' of Lagrangian fibers is being described. This coordinate
choice is explained in further detail in \cite{AKV02}. In this note,
the relevant coordinates are the $y_{2}=1$ patch. In addition, to
achieve later agreement with conventions from knot contact homology,
the following change of coordinates will be used: 
\begin{align*}
Q & :=e^{t}, & x & :=-y_{3}/Q, & p:= & -y_{4}/Q.
\end{align*}
Then, in this patch and coordinate system, the mirror manifold is
\begin{equation}
\hat{X}=\left\{ uv=1-Qx-Qp+Qxp\right\} .\label{eq:mirror-of-X}
\end{equation}

\begin{example*}
[Open Gromov-Witten invariants via mirror symmetry] Consider Lagrangian
fibers $L$ of type I. These Lagrangians have topology $S^{1}\times\mathbb{R}^{2}$,
and intersect the base $\mathbb{P}^{1}$ along the $S^{1}$. In the
mirror, the moduli space of such Lagrangian fibers is the Riemann
surface $S\subset\hat{X}$ defined by setting $uv=0$: 
\[
S=\left\{ 1-Qx-Qp+Qxp=0\right\} .
\]
Mirror symmetry equates the periods of certain differential forms
on $\hat{X}$ to generating functions for open Gromov-Witten invariants.
In this case, the prediction of mirror symmetry is that (up to constant
factors in $x$) 
\begin{equation}
\int\lambda=\sum_{d,w}K_{d,w}Q^{d}x^{w},\label{eq:genus-0-mirror-symmetry}
\end{equation}
where $\lambda:=-\log p\frac{dx}{x}$ is a one-form along $S$ defined
by solving $1-Qx-Qp+Qxp=0$ for $p$, and $K_{d,w}$ is the genus-0,
degree-$d$, winding $w$ open Gromov-Witten invariant with boundary
on $L$. In terms of $x$ and $Q$, 
\begin{align*}
-\log p\left(x;Q\right) & =\log\left(Q\right)+\log\left(\frac{1-x}{1-Qx}\right)\\
 & =\log\left(Q\right)+\sum_{n=1}^{\infty}\frac{1}{n}\left(-1+Q^{n}\right)x^{n},
\end{align*}
so (\ref{eq:genus-0-mirror-symmetry}) asserts that 
\[
\sum_{d,w}K_{d,w}Q^{d}x^{w}=\sum_{n=1}^{\infty}\frac{1}{n^{2}}\left(-1+Q^{n}\right)x^{n}.
\]
Hence 
\[
K_{d,w}=\begin{cases}
-\frac{1}{w^{2}}, & d=0;\\
\frac{1}{d^{2}}, & d=w;\\
0 & \mbox{otherwise.}
\end{cases}
\]
Note that in this case, $K_{d,w}$ has also been computed directly
using localization in \cite{KL01}, and these results agree in framing
0. 
\end{example*}
Open Gromov-Witten invariants are also conjectured to satisfy certain
integrality requirements \cite{LMV00,AV00}. For genus-zero, one-boundary-component
invariants, the requirement is that 
\begin{equation}
K_{d,w}=\sum_{n|d\mbox{ and }n|w}\frac{1}{n^{2}}N_{d/n,w/n},\label{eq:integrality}
\end{equation}
where $N_{d,w}\in\mathbb{Z}$. In terms of the generating function
for $K_{d,w}$, this is 
\[
\sum_{d,w}K_{d,w}Q^{d}x^{w}=\sum_{n>0}\sum_{d,w}\frac{1}{n^{2}}N_{d,w}.
\]
In the previous example, $N_{0,1}=-1$, $N_{1,1}=1$, and $N_{d,w}=0$
for all other $d$, $w$.

\subsection{Knots and the Conifold Transition\label{sub:conifold-trans}}

The Lagrangian submanifolds described by (\ref{eq:sLag-fibers}) have
a very specific geometry. A natural question is, what are the open
Gromov-Witten invariants associated to Lagrangians with a different
geometry? One way of obtaining such Lagrangians is through the conifold
transition \cite{OV00,DSV13}. The manifold $X=\mathcal{O}_{\mathbb{P}^{1}}(-1,-1)$
can also be obtained as the resolution of the conifold singularity
in $\mathbb{C}^{4}$---it is given by the equations 
\begin{align*}
xz-yw & =0,\\
x\lambda & =w\rho,\\
y\lambda & =z\rho
\end{align*}
where $\left((x,y,z,w),\left[\lambda:\rho\right]\right)\in\mathbb{C}^{4}\times\mathbb{P}^{1}$
. The conifold singularity $xz-yw=0$ is the limit of the hypersurface
\[
Y_{\mu}:=\left\{ xz-yw=\mu\right\} \subset\mathbb{C}^{4}
\]
as $\mu\rightarrow0$. For $\mu\neq0$, $Y_{\mu}$ is symplectomorphic
to the cotangent bundle $T^{*}S^{3}$. The zero section $S_{\mu}\cong S^{3}\subset Y_{\mu}$
is the fixed locus of the antiholomorphic involution $(x,y,z,w)\mapsto(\overline{z},-\overline{w},\overline{x},-\overline{y})$,
and is described by the equation $\left|x\right|^{2}+\left|y\right|^{2}=\mu$.
Away from the zero sections, the conifold transition gives a symplectomorphism
between $T^{*}S^{3}$ and $X$ \cite{DSV13}. Thus, a Lagrangian submanifold
of $T^{*}S^{3}$ which does not intersect the zero section $S^{3}$
is symplectomorphic to a Lagrangian submanifold of $X$ which also
does not intersect the zero section. 

\begin{figure}
\begin{center}
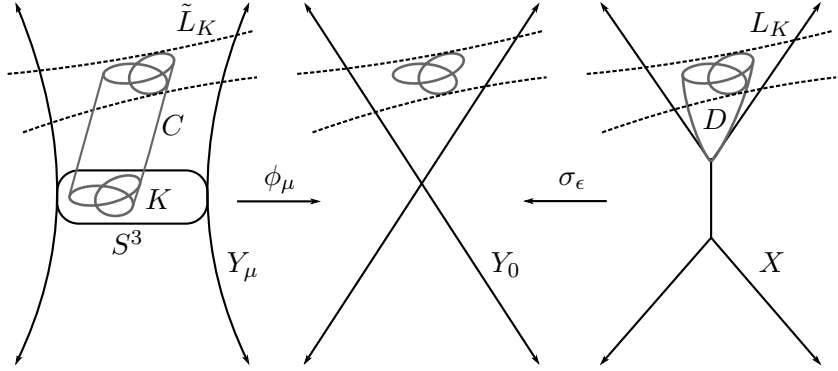
\end{center}

\caption{\label{fig:conifold-transition}The conifold transition and knots.}

\begin{minipage}[t]{1\columnwidth}%
The Lagrangian $\tilde{L}_{K}\subset Y_{\mu}\cong T^{*}S^{3}$ is
constructed by shifting the conormal bundle of a knot $K\subset S^{3}$
off of the zero section. This lift introduces a holomorphic cylinder
$C$ connecting the knot on $S^{3}$ to its image in $\tilde{L}_{K}$.
$Y_{0}$ is the conifold singularity $xz-yw=0$ in $\mathbb{C}^{4}$.
The map $\phi_{\mu}:Y_{\mu}\rightarrow Y_{0}$ is a symplectomorphism
away from the zero section, so $\phi_{\mu}\left(\tilde{L}_{K}\right)$
is a Lagrangian submanifold of $Y_{0}$. $X\cong\mathcal{O}_{\mathbb{P}^{1}}\left(-1,-1\right)$
is the small resolution of the conifold singularity, and $\sigma_{\epsilon}:X\rightarrow Y_{0}$
is the corresponding natural map. In fact, there are a family of such
symplectomorphisms, where $\epsilon$ parametrizes the symplectic
form on the zero section $\mathbb{P}^{1}\subset X$. Hence, $L_{K}:=\sigma_{\epsilon}^{-1}\circ\phi_{\mu}\left(\tilde{L}_{K}\right)$
is a Lagrangian submanifold of $X$. The holomorphic disk $D$ is
the image of $C$ under the conifold transition. %
\end{minipage}
\end{figure}

Knots $K\subset S^{3}$ are a source of Lagrangian submanifolds in
$T^{*}S^{3}$: the conormal bundle $N_{K}^{*}\subset T^{*}S^{3}$
is Lagrangian with topology $S^{1}\times\mathbb{R}^{2}$. In order
to obtain a Lagrangian submanifold of $X$ from $N_{K}^{*}$, $N_{K}^{*}$
must first be moved off of the zero section. This can be done by choosing
a lift of $K$, as described in \cite{DSV13}. The image of the shifted
conormal bundle will be a Lagrangian submanifold $L_{K}\subset X$,
as depicted in Figure~\ref{fig:conifold-transition}. 

\subsection{Open Gromov-Witten Invariants and the Augmentation Polynomial\label{sub:open-GW-invts-from-aug-poly}}

For torus knots, the corresponding open Gromov-Witten invariants of
$(X,L_{K})$ have been computed directly using localization in \cite{DSV13},
and using mirror symmetry mirror symmetry in \cite{BEM12,JKS14}.
However, the mirror symmetry approach of \cite{BEM12} does not readily
generalize to non-toric knots. In this subsection, a recent conjecture
of Aganagic and Vafa \cite{AV12} is applied to compute open Gromov-Witten
invariants via an analogous mirror symmetry computation. This approach
uses the augmentation polynomial from Legendrian contact homology.
In contrast to localization and the method of \cite{BEM12}, this
method can be applied for any knot whose augmentation polynomial is
known---including non-toric knots. 

The Aganagic-Vafa conjecture can be motivated by observing that the
mirror to $\hat{X}$ given in (\ref{eq:mirror-of-X}) can be written
as 
\[
\hat{X}=\left\{ uv=A(x,p;Q)\right\} ,
\]
where $A(x,p;Q)=1-Qx-Qp+Qxp$ is the augmentation polynomial of the
unknot in Legendrian contact homology \cite{EENS13,Ng12}. The moduli
of Lagrangian submanifolds with topology $S^{1}\times\mathbb{R}^{2}$
was described by the zero locus 
\[
\left\{ A(x,p;Q)=0\right\} \subset\hat{X}.
\]
Moreover, the image of a shifted conormal bundle to the unknot under
the conifold transition described in Section~\ref{sub:conifold-trans}
is exactly one of these Lagrangian fibers \cite{OV00,GJKS15}. Hence,
one might speculate that the moduli of Lagrangian submanifolds $L_{K}\subset X$
obtained from the conormal bundles of other knots $K\subset S^{3}$
are given by the Riemann surface 
\[
\left\{ A_{K}(x,p;Q)=0\right\} ,
\]
where $A_{K}(x,p;Q)$ is the augmentation polynomial of the knot $K$. 

Aganagic and Vafa then conjecture that for each choice of knot, there
is a corresponding mirror $\hat{X}_{K}$ describing the moduli space
of Lagrangian fibers with geometry $L_{K}$ along the singular locus,
and that $\hat{X}_{K}$ is described by 
\[
\hat{X}_{K}:=\left\{ uv=A_{K}(x,p;Q)\right\} .
\]
In addition to the apparent coincidence between the unknot's augmentation
polynomial and the mirror $\hat{X}$, there are also physical arguments
for this conjecture coming from the connections between topological
string theory, Chern-Simons theory, and the HOMFLY polynomial \cite{AV12,AENV14,JKS14,NW07}.
The augmentation polynomial $A_{K}$ can be identified with the classical
limit of the $Q$-deformed quantum A-polynomial, which satisfies an
elimination condition on the Chern-Simons partition function in the
presence of a Wilson loop coming from $K$. This Chern-Simons partition
function can in turn be obtained from the colored HOMFLY polynomials
of $K$, and is identified with the open Gromov-Witten generating
function under the conifold transition. 

For the purposes of computing genus-zero open Gromov-Witten invariants,
the application of this conjecture is straightforward: as in Section~\ref{sub:standard-mirror},
the open Gromov-Witten generating function is equated with the integral
of a differential form: 
\[
\sum_{d,w}K_{d,w}Q^{d}x^{w}=\int-\log p(x;Q)\frac{dx}{x}.
\]
However, $p(x;Q)$ is now determined by the equation 
\[
A_{K}(x,p;Q)=0.
\]
The main difficulty is that $A_{K}(x,p;Q)$ may be a high-order polynomial
in $p$, so it is not always feasible to find analytic solutions for
$p$. Even for torus knots, this can be an obstacle: for an $(r,s)$
torus knot, 
\begin{equation}
\begin{aligned}\deg p & =\left(\begin{matrix}r+s\\
s
\end{matrix}\right), & \deg x & =\frac{\deg p}{r+s}\end{aligned}
\label{eq:degrees-of-A-torus-knots}
\end{equation}
are the maximum degrees \cite{JKS14}. 

Fortunately, to compute $K_{d,w}$ for a given $d$ and $w$, only
a series solution for $p$ is needed. Suppose that 
\begin{equation}
p(x;Q)=\exp\left(-\sum_{n=0}^{\infty}W_{n}(Q)x^{n}\right),\label{eq:p-series}
\end{equation}
where $W_{n}(Q)$ is a polynomial in $Q$ for $n>0$, and $W_{0}(Q)$
determines the overall scaling of $p$. (Note that in all considered
examples, $W_{0}(Q)=\log(Q)$, i.e., $p(x;Q)\approx1/Q+\cdots$).
Then, substitute this expression into 
\[
A_{K}\left(x,p(x;Q);Q\right)=0.
\]
The resulting expression will be a series in $x$, which can be solved
by recursively finding the coefficients $W_{n}(Q)$---in general,
the coefficient of $x^{n}$ will be a polynomial function of $\left\{ W_{k}(Q)\right\} _{k\leq n}$.
The coefficients $W_{n}(Q)$ are related to the $K_{d,w}$ by 
\[
W_{n}(Q)=\sum_{d}nK_{d,n}Q^{d}.
\]
For completeness, the integer invariants $N_{d,w}$ can be similarly
determined from the $K_{d,w}$: (\ref{eq:integrality}) can be re-written
as 
\[
N_{d,w}=K_{d,w}-\sum_{\substack{n|d\mbox{ and }n|w\\
n>1
}
}\frac{1}{n^{2}}N_{d/n,w/n},
\]
so by starting with $N_{0,1}=K_{0,1}$, successive $N_{d,w}$ can
be solved for. The following section uses this method to compute $K_{d,w}$
and $N_{d,w}$ for two non-toric knots.

\section{Non-Toric Examples\label{sec:Non-Toric-Examples}}

As compared to Atiyah-Bott localization, one advantage of computing
open Gromov-Witten invariants from the augmentation polynomial is
that this technique does not require that the Lagrangian $L$ is fixed
by a torus action. Such Lagrangians can be obtained from the conormal
bundles of non-toric knots. This section performs the computation
of Section~\ref{sub:open-GW-invts-from-aug-poly} for two non-toric
knots: the $4_{1}$ (figure-eight) knot and the $5_{2}$ (three-twist)
knot. In contrast to torus knots, these knots cannot be expressed
as the link of the singularity $x^{r}-y^{s}=0$ in $S^{3}$ for any
$r,s\in\mathbb{Z}$. For non-toric knots, it is not currently known
how to directly compute the open Gromov-Witten invariants $K_{d,w}$
using localization. However, the invariants computed in the following
examples have been checked to satisfy the integrality condition (\ref{eq:integrality})
for all $w\leq8$ and arbitrary $d$.

\subsection{The $4_{1}$ (figure-eight) knot}

The $4_{1}$ knot is the unique knot with crossing number 4, and is
a twist knot obtained from two half-twists. According to \cite{AV12,GJKS15},
the augmentation polynomial of the $4_{1}$ knot in framing 0 is 
\begin{align*}
A_{K}\left(x,p,Q\right) & =p^{2}-Qp^{3}+\left(Q^{3}p^{5}-2Q^{3}p^{4}+2Qp-Q\right)x\\
 & +\left(-Q^{5}p^{5}+2Q^{4}p^{4}-2Q^{3}p+Q^{2}\right)x^{2}\\
 & +\left(Q^{5}p^{3}-Q^{5}p^{2}\right)x^{3}.
\end{align*}
Rescaling $x$ by $x\mapsto Qx$, one can obtain a series solution
for $-\log p$ using the method of section~\ref{sub:open-GW-invts-from-aug-poly}.
The first few terms are 
\begin{align*}
-\log p & =\log\left(Q\right)+\left(Q^{3}-2Q^{2}+2Q-1\right)x\\
 & +\left(\frac{5Q^{6}}{2}-8Q^{5}+9Q^{4}-9Q^{2}+8Q-\frac{5}{2}\right)x^{2}+\cdots
\end{align*}
The coefficients of this series solution determine the open Gromov-Witten
invariants $K_{d,w}$ and corresponding integer invariants $N_{d,w}$.
For $w\leq4$ $0\leq d\leq3$, these invariants are listed in Table~\ref{tab:41-knot}.
The integrality condition $N_{d,w}\in\mathbb{Z}$ has also been verified
for all $w\leq8$ and arbitrary $d$. Note that for $w\leq8$, $K_{d,w}=N_{d,w}=0$
for any $d>24$. 
\begin{table}
\caption{\label{tab:41-knot}$K_{d,w}$ and $N_{d,w}$ for the $4_{1}$ knot.}

{\renewcommand{\arraystretch}{1.25}

\hfill{}%
\begin{tabular}{r|rrrr}
$K_{d,w}$ & $K_{d,1}$ & $K_{d,2}$ & $K_{d,3}$ & $K_{d,4}$\tabularnewline
\hline 
$K_{0,w}$ & $-1$  & $-\frac{5}{4}$  & $-\frac{28}{9}$  & $-\frac{165}{16}$ \tabularnewline
$K_{1,3}$ & $2$  & $4$  & $14$  & $60$ \tabularnewline
$K_{2,w}$ & $-2$  & $-\frac{9}{2}$  & $-25$  & $-147$ \tabularnewline
$K_{3,w}$ & $1$  & $0$  & $\frac{173}{9}$  & $186$ \tabularnewline
\end{tabular}\hfill{}%
\begin{tabular}{r|rrrr}
$N_{d,w}$ & $N_{d,1}$ & $N_{d,2}$ & $N_{d,3}$ & $N_{d,4}$\tabularnewline
\hline 
$N_{0,w}$ & $-1$  & $-1$  & $-3$  & $-10$ \tabularnewline
$N_{1,3}$ & $2$  & $4$  & $14$  & $60$ \tabularnewline
$N_{2,w}$ & $-2$  & $-5$  & $-25$  & $-148$ \tabularnewline
$N_{3,w}$ & $1$  & $0$  & $19$  & $186$ \tabularnewline
\end{tabular}\hfill{}}
\end{table}

\subsection{The $5_{2}$ (three-twist) knot}

The $5_{2}$ knot is a twist knot obtained from three half-twists,
and is one of two knots with crossing number 5 (the other being the
$(5,2)$ torus knot). The augmentation polynomial for the $5_{2}$
knot is 
\begin{align*}
A_{K}(x,p,Q) & =Q^{2}p^{8}-Qp^{7}+x^{4}\left(-p+1\right)\\
 & +\left(-Q^{3}p^{6}+2Q^{2}p^{5}-Qp^{4}-2Qp^{3}+3Qp^{2}-3p^{2}+4p-2\right)x^{3}\\
 & +\left(Q^{4}p^{8}-3Q^{3}p^{7}-4Q^{3}p^{6}+5Q^{2}p^{6}+3Q^{2}p^{5}\right.\\
 & \left.+6Q^{2}p^{4}-3Qp^{5}-4Qp^{4}+3Qp^{3}-4Qp^{2}-3p^{3}+5p^{2}-3p+1\right)x^{2}\\
 & +\left(-2Q^{3}p^{8}+4Q^{2}p^{7}+3Q^{2}p^{6}-3Qp^{6}-2Qp^{5}-p^{4}+2p^{3}-p^{2}\right)x.
\end{align*}
Rescaling $x$ by $x\mapsto x/Q$, the method of \ref{sub:open-GW-invts-from-aug-poly}
gives a series solution for $-\log p$. The first few terms are 
\begin{align*}
-\log p & =\log\left(Q\right)+\left(-Q^{4}+2Q^{3}-Q\right)x\\
 & +\left(\frac{11Q^{8}}{2}-20Q^{7}+23Q^{6}-8Q^{5}+2Q^{4}-4Q^{3}+\frac{3Q^{2}}{2}\right)x^{2}+\cdots
\end{align*}
The corresponding $K_{d,w}$ and $N_{d,w}$ are obtained from the
coefficients of this series solution. These invariants are listed
in Table~\ref{tab:52-knot}. As for the $4_{1}$ knot, the integrality
condition $N_{d,w}\in\mathbb{Z}$ has also been checked for all $w\leq8$,
and it was again found that $K_{d,w}=N_{d,w}=0$ for any $w\leq8$
and $d>24$. 
\begin{table}
\caption{\label{tab:52-knot}$K_{d,w}$ and $N_{d,w}$ for the $5_{2}$ knot.}

{\renewcommand{\arraystretch}{1.25}

\hfill{}%
\begin{tabular}{r|rrrr}
$K_{d,w}$ & $K_{d,1}$ & $K_{d,2}$ & $K_{d,3}$ & $K_{d,4}$\tabularnewline
\hline 
$K_{0,w}$ & $-1$  & $\frac{3}{4}$  & $-\frac{10}{9}$  & $\frac{35}{16}$ \tabularnewline
$K_{1,3}$ & $0$  & $-2$  & $4$  & $-12$ \tabularnewline
$K_{2,w}$ & $2$  & $1$  & $0$  & $\frac{27}{2}$ \tabularnewline
$K_{3,w}$ & $-1$  & $-4$  & $-12$  & $8$ \tabularnewline
\end{tabular}\hfill{}%
\begin{tabular}{r|rrrr}
$N_{d,w}$ & $N_{d,1}$ & $N_{d,2}$ & $N_{d,3}$ & $N_{d,4}$\tabularnewline
\hline 
$N_{0,w}$ & $-1$  & $1$  & $-1$  & $2$ \tabularnewline
$N_{1,3}$ & $0$  & $-2$  & $4$  & $-12$ \tabularnewline
$N_{2,w}$ & $2$  & $1$  & $0$  & $14$ \tabularnewline
$N_{3,w}$ & $-1$  & $-4$  & $-12$  & $8$ \tabularnewline
\end{tabular}\hfill{}}
\end{table}

\section{Recovering the Augmentation Polynomial\label{sec:Recovering-the-aug-poly}}

The augmentation polynomial conjecturally contains all of the open
Gromov-Witten invariants $K_{d,w}$ for a given Lagrangian brane $L_{K}$.
As seen in the previous section, open Gromov-Witten invariants can
be extracted from the augmentation polynomial. For torus knots, the
genus zero open Gromov-Witten invariants have also been computed directly
via localization in \cite{MahowaldThesis,DSV13}. Explicitly, 
\begin{equation}
K_{d,w}^{(r,s)}=\left(-1\right)^{d+1}\left(\frac{\prod_{k=0}^{d-1}\left(wr-k\right)\left(ws-k\right)}{wd!\prod_{k=1}^{d}\left(wr+ws-k\right)}\right)\left(\frac{\prod_{k=1}^{ws-1}\left(r+s-\frac{k}{w}\right)}{w\prod_{k=0}^{ws-1}\left(s-\frac{k}{w}\right)}\right),\label{eq:Kdw-r-s}
\end{equation}
where $r>s>0$ are coprime, and $ws\geq d\geq0$. For $d>ws$, $K_{d,w}^{(r,s)}=0$.
This section describes a method for obtaining the augmentation polynomial
$A_{K}$ when $K$ is a torus knot, and implements this method for
two examples (the unknot, and the $(3,2)$ torus knot). 

The idea is to use (\ref{eq:Kdw-r-s}) and (\ref{eq:p-series}) to
obtain a system of linear equations on the coefficients of the augmentation
polynomial. Write 
\begin{equation}
A_{K}(x,p;Q)=\sum_{j,k}c_{jk}x^{j}\left(p(x;Q)\right)^{k},\label{eq:Augpoly-series}
\end{equation}
where $c_{jk}$ is a rational polynomial in $Q$. Recall that for
torus knots, the degrees of $x$ and $p$ are given by (\ref{eq:degrees-of-A-torus-knots}).
Let 
\[
p(x;Q)=\frac{1}{Q}\exp\left(-\sum_{d,w}wK_{d,w}Q^{d}x^{w}\right)
\]
and $W_{n}:=\sum_{d\geq0}nK_{d,n}Q^{d}$. Then the coefficient of
$x^{n}$ in a series expansion of $\left(p(x;Q)\right)^{k}$ is $P_{n}(k)/Q^{k}$,
where 
\[
P_{n}(k):=\sum_{\substack{i_{1}+2i_{2}+\cdots+ni_{n}=n\\
i_{j}\geq0
}
}\left[\prod_{j=1}^{n}\frac{(-1)^{i_{j}}}{i_{j}!}\left(kW_{j}\right)^{i_{j}}\right].
\]
($P_{0}(k)=1$). Substituting this expression into (\ref{eq:Augpoly-series})
gives a power series in $x$: 
\begin{align*}
A_{K}(x,p;Q) & =\sum_{j,k}c_{jk}\left(\frac{x^{j}}{Q^{k}}\right)\left(\sum_{n\geq0}P_{n}(k)x^{n}\right)\\
 & =\sum_{n\geq0}\left(\sum_{j=0}^{n}\sum_{k\geq0}\frac{P_{n-j}(k)}{Q^{k}}c_{jk}\right)x^{n}.
\end{align*}
In order for $p(x;Q)$ to be a solution of $A_{K}(x,p;Q)=0$, the
coefficient of $x^{n}$ for all $n$ in the above expression must
vanish. This gives a collection of linear equations in the $c_{jk}$:
\begin{equation}
\sum_{j=0}^{n}\sum_{k\geq0}\frac{P_{n-j}(k)}{Q^{k}}c_{jk}=0,\label{eq:cjk}
\end{equation}
which can be solved to determine the $c_{jk}$ up to an overall rescaling
of the augmentation polynomial. (Note that such rescalings do not
affect the solutions $p(x;Q)$, and hence do not affect the Gromov-Witten
invariants). The following two examples implement this procedure for
the unknot in framing 0 and the (3,2) torus knot in framing 6. For
both knots, this method recovers the expected augmentation polynomial.
However, for the (3,2) knot, due to computational complexity, some
simplifying assumptions about the values of the $c_{jk}$ are made.

\subsection{The unknot}

For the unknot, $\deg x=1$, $\deg p=1$, so there are four coefficients
to solve for: $c_{00}$, $c_{01}$, $c_{10}$, and $c_{11}$. Here,
the $W_{n}$ have a simple expression: 
\[
W_{n}=\frac{1}{n}\left(Q^{n}-1\right).
\]
The first four linear equations from (\ref{eq:cjk}) are: 
\begin{align*}
c_{00}+\frac{1}{Q}c_{01} & =0, & (n & =0)\\
\frac{1}{Q}(1-Q)c_{01}+c_{10}+\frac{1}{Q}c_{11} & =0, & (n & =1)\\
\frac{1}{Q}(1-Q)c_{01}+\frac{1}{Q}(1-Q)c_{11} & =0. & (n & =2,3)
\end{align*}
(The $n=2$ and $n=3$ equations both simplify to the same expression.)
With $c_{00}$ as the free variable, these equations become 
\begin{align*}
c_{01} & =-Qc_{00,} & c_{10} & =-Qc_{00}, & c_{11} & =Qc_{00}.
\end{align*}
Normalizing to $c_{00}=1$ yields the expected augmentation polynomial
of the unknot in framing 0: 
\[
A_{K}(x,p;Q)=1-Qx-Qp+Qxp.
\]

\subsection{The $(3,2)$ torus knot}

The augmentation polynomial of the (3,2) torus knot is 
\begin{align*}
A_{K}\left(x,p;Q\right) & =1-Qp+\left(Q^{5}p^{3}-Q^{5}p^{4}+2Q^{5}p^{5}-2Q^{6}p^{5}-Q^{6}p^{6}+Q^{7}p^{7}\right)x\\
 & +\left(Q^{10}p^{10}-Q^{10}p^{9}\right)x^{2}.
\end{align*}
For the $(3,2)$ torus knot, $\deg p=10$ and $\deg x=2$. So, $c_{jk}=0$
for all $j>2$ and $k>10$. Note that in general, one would need to
consider at least $\left(\deg p+1\right)\left(\deg x+1\right)$ equations
to determine $c_{jk}$. Due to the increasing complexity of the equations
involved, in this example the following simplifying assumptions will
be made: $c_{0k}=0$ for $k>1$, $c_{10}=c_{11}=c_{12}=0$, $c_{18}=c_{19}=c_{1,10}=0$,
and $c_{2k}=0$ for $k<9$. Thus, the remaining nine ``unknown''
coefficients are $c_{00}$, $c_{01}$, $c_{13}$, $c_{14}$, $c_{15}$,
$c_{16}$, $c_{17}$, $c_{29}$, and $c_{2,10}$. The first three
equations from (\ref{eq:cjk}) are: 
\begin{align*}
c_{00}+\frac{1}{Q}c_{01} & =0,\\
\left(2Q^{-1}-3+Q\right)c_{01}+Q^{-3}c_{13}+Q^{-4}c_{14}+Q^{-5}c_{15}+Q^{-6}c_{16}+Q^{-7}c_{17} & =0,\\
\left(23Q^{-1}-62+59Q-23Q^{2}+3Q^{3}\right)c_{01}+\left(3Q^{-1}-9Q^{-2}+6Q^{-3}\right)c_{13}\\
+\left(4Q^{-2}-12Q^{-3}+8Q^{-4}\right)c_{14}+\left(5Q^{-3}-15Q^{-4}+10Q^{-5}\right)c_{15}\\
+\left(6Q^{-4}-18Q^{-5}+12Q^{-6}\right)c_{16}+\left(7Q^{-5}-21Q^{-6}+14Q^{-7}\right)c_{17}\\
+Q^{-9}c_{29}+Q^{-10}c_{2,10} & =0.
\end{align*}
(For brevity, the remaining equations are omitted). Solving for the
$c_{ij}$ (with $c_{00}$ as the free variable), one finds 
\begin{align*}
c_{01} & =-Qc_{00}, & c_{13} & =Q^{5}c_{00}, & c_{14} & =-Q^{5}c_{00}, & c_{15} & =\left(2Q^{5}-2Q^{6}\right)c_{00},\\
c_{16} & =-Q^{6}c_{00}, & c_{17} & =Q^{7}c_{00}, & c_{29} & =-Q^{10}c_{00}, & c_{2,10} & =Q^{10}c_{00}.
\end{align*}
 By normalizing to $c_{00}=1$, the expected coefficients of the augmentation
polynomial,
\begin{align*}
c_{01} & =-Q, & c_{13} & =Q^{5}, & c_{14} & =-Q^{5}, & c_{15} & =2Q^{5}-2Q^{6},\\
c_{16} & =-Q^{6}, & c_{17} & =Q^{7}, & c_{29} & =-Q^{10}, & c_{2,10} & =Q^{10},
\end{align*}
 are obtained. 

\bibliographystyle{plain}
\bibliography{bibliography}

\begin{thebibliography}{10}

\bibitem{AENV14}
M.~Aganagic, T.~Ekholm, L.~Ng, and C.~Vafa.
\newblock Topological strings, {D}-model, and knot contact homology.
\newblock {\em Adv. Theor. Math. Phys.}, 18(4):827--956, 2014.
\newblock \arxiv{1304.5778}{~[hep-th]}.

\bibitem{AKV02}
M.~Aganagic, A.~Klemm, and C.~Vafa.
\newblock Disk instantons, mirror symmetry and the duality web.
\newblock {\em Z. Naturforsch. A}, 57(1-2):1--28, 2002.
\newblock \arxiv{hep-th/0105045}{~[hep-th]}.

\bibitem{AV00}
M.~Aganagic and C.~Vafa.
\newblock Mirror symmetry, {D}-branes and counting holomorphic discs.
\newblock 2000.
\newblock \arxiv{hep-th/0012041}{}.

\bibitem{AV12}
M.~Aganagic and C.~Vafa.
\newblock Large {N} duality, mirror symmetry, and a {Q}-deformed {A}-polynomial
  for knots.
\newblock 2012.
\newblock \arxiv{1204.4709}{~[hep-th]}.

\bibitem{BEM12}
A.~Brini, B.~Eynard, and M.~Mari{\~n}o.
\newblock Torus knots and mirror symmetry.
\newblock {\em Ann. Henri Poincar\'e}, 13(8):1873--1910, 2012.
\newblock \arxiv{1105.2012}{~[hep-th]}.

\bibitem{COGP91}
P.~Candelas, X.~C. de~la Ossa, P.~S. Green, and L.~Parkes.
\newblock A pair of {C}alabi-{Y}au manifolds as an exactly soluble
  superconformal theory.
\newblock {\em Nuclear Phys. B}, 359(1):21--74, 1991.

\bibitem{CKYZ99}
T.-M. Chiang, A.~Klemm, S.-T. Yau, and E.~Zaslow.
\newblock Local mirror symmetry: calculations and interpretations.
\newblock {\em Adv. Theor. Math. Phys.}, 3(3):495--565, 1999.

\bibitem{DSV13}
D.-E. Diaconescu, V.~Shende, and C.~Vafa.
\newblock Large {$N$} duality, {L}agrangian cycles, and algebraic knots.
\newblock {\em Comm. Math. Phys.}, 319(3):813--863, 2013.
\newblock \arxiv{1111.6533v1}{~[hep-th]}.

\bibitem{EENS13}
T.~Ekholm, J.~Etnyre, L.~Ng, and M.~Sullivan.
\newblock Knot contact homology.
\newblock {\em Geom. Topol.}, 17(2):975--1112, 2013.
\newblock \arxiv{1109.1542}{~[math.SG]}.

\bibitem{GV99}
R.~Gopakumar and C.~Vafa.
\newblock On the gauge theory/geometry correspondence.
\newblock {\em Adv. Theor. Math. Phys.}, 3(5):1415--1443, 1999.
\newblock \arxiv{hep-th/9811131}{}.

\bibitem{GP99}
T.~Graber and R.~Pandharipande.
\newblock Localization of virtual classes.
\newblock {\em Invent. Math.}, 135(2):487--518, 1999.
\newblock \arxiv{alg-geom/9708001}{}.

\bibitem{GJKS15}
J.~Gu, H.~Jockers, A.~Klemm, and M.~Soroush.
\newblock Knot invariants from topological recursion on augmentation varieties.
\newblock {\em Comm. Math. Phys.}, 336(2):987--1051, 2015.
\newblock \arxiv{1401.5095v3}{~[hep-th]}.

\bibitem{HV00}
K.~Hori and C.~Vafa.
\newblock Mirror symmetry.
\newblock 2000.
\newblock \arxiv{hep-th/0002222}{}.

\bibitem{JKS14}
H.~Jockers, A.~Klemm, and M.~Soroush.
\newblock Torus knots and the topological vertex.
\newblock {\em Lett. Math. Phys.}, 104(8):953--989, 2014.
\newblock \arxiv{1212.0321}{~[hep-th]}.

\bibitem{KL01}
S.~Katz and C.-C.~M. Liu.
\newblock Enumerative geometry of stable maps with {L}agrangian boundary
  conditions and multiple covers of the disc.
\newblock {\em Adv. Theor. Math. Phys.}, 5(1):1--49, 2001.
\newblock \arxiv{math/0103074}{~[math.AG]}.

\bibitem{KZ99}
A.~Klemm and E.~Zaslow.
\newblock Local mirror symmetry at higher genus.
\newblock 1999.
\newblock \arxiv{hep-th/9906046}{}.

\bibitem{Kontsevich95}
M.~Kontsevich.
\newblock Enumeration of rational curves via torus actions.
\newblock In {\em The moduli space of curves ({T}exel {I}sland, 1994)}, volume
  129 of {\em Progr. Math.}, pages 335--368. Birkh\"auser Boston, Boston, MA,
  1995.
\newblock \arxiv{hep-th/9405035v2}{}.

\bibitem{LMV00}
J.~Labastida, M.~Mari{\~n}o, and C.~Vafa.
\newblock Knots, links and branes at large {$N$}.
\newblock {\em J. High Energy Phys.}, (11):Paper 7, 42, 2000.
\newblock \arxiv{hep-th/0010102}{}.

\bibitem{MahowaldThesis}
M.~Mahowald.
\newblock {\em Knots and {G}amma classes in open topological string theory}.
\newblock PhD thesis, Northwestern University, Evanston, Illinois, June 2016.

\bibitem{NW07}
A.~Neitzke and J.~Walcher.
\newblock background independence and the open topological string wavefunction.
\newblock 2007.
\newblock \arxiv{0709.2390}{~[hep-th]}.

\bibitem{Ng12}
L.~Ng.
\newblock A topological introduction to knot contact homology.
\newblock 2012.
\newblock \arxiv{1210.4803}{~[math.GT]}.

\bibitem{OV00}
H.~Ooguri and C.~Vafa.
\newblock Knot invariants and topological strings.
\newblock {\em Nuclear Phys. B}, 577(3):419--438, 2000.
\newblock \arxiv{hep-th/9912123}{}.

\bibitem{SYZ96}
A.~Strominger, S.-T. Yau, and E.~Zaslow.
\newblock Mirror symmetry is {$T$}-duality.
\newblock {\em Nuclear Phys. B}, 479(1-2):243--259, 1996.

\bibitem{Witten89}
E.~Witten.
\newblock Quantum field theory and the {J}ones polynomial.
\newblock {\em Comm. Math. Phys.}, 121(3):351--399, 1989.

\bibitem{Witten92}
E.~Witten.
\newblock Mirror manifolds and topological field theory.
\newblock In {\em Essays on mirror manifolds}, pages 120--158. Int. Press, Hong
  Kong, 1992.
\newblock \arxiv{hep-th/9112056}{}.

\end{thebibliography}

\end{document}